\documentclass[conference]{IEEEtran}
\IEEEoverridecommandlockouts
\usepackage{cite}
\usepackage{amsmath,amssymb,amsfonts}
\usepackage{graphicx}
\usepackage{textcomp}
\usepackage{xcolor}
\usepackage{enumitem}
\PassOptionsToPackage{bookmarks=false}{hyperref}
\usepackage{eso-pic}

\usepackage{booktabs}
\usepackage{multirow}

\usepackage{stfloats}

\usepackage{url}
\usepackage{algorithm}
\usepackage{algpseudocode}

\def\BibTeX{{\rm B\kern-.05em{\sc i\kern-.025em b}\kern-.08em
    T\kern-.1667em\lower.7ex\hbox{E}\kern-.125emX}}

\begin{document}

\AddToShipoutPictureFG*{%
  \AtPageUpperLeft{%
    \raisebox{-0.8cm}{%
      \hspace{4.5cm}%
      \large\textbf{Presented at IEEE MILCOM 2026 (National Capital Region, USA)}
    }%
  }%
}

\title{The Impact of Demand Forecasting on Delay and Jitter in DVB-Based Beam-Hopping LEO Networks}

\author{\IEEEauthorblockN{Yekta Demirci\IEEEauthorrefmark{1},
Guillaume Mantelet\IEEEauthorrefmark{2},
Stéphane Martel\IEEEauthorrefmark{2},
Jean-François Frigon\IEEEauthorrefmark{1},
Gunes Karabulut Kurt\IEEEauthorrefmark{1}}
\IEEEauthorblockA{\IEEEauthorrefmark{1}Poly-Grames Research Center, Department of Electrical Engineering, Polytechnique Montréal, QC, Canada}
\IEEEauthorblockA{\IEEEauthorrefmark{2} Satellite Systems, MDA Space, Canada}
}

\maketitle

\begin{abstract}
In LEO satellite networks utilizing beam hopping (BH), resource allocation plans must be committed well in advance. This inherent operational delay necessitates predicting future user demand during the planning phase. Such predictive agility is particularly crucial for military applications, where unpredictable tactical environments demand low-latency, resilient communication links. However, existing forecasting models are typically evaluated based on standalone accuracy, ignoring their cross-layer impact on overall network performance. To address this gap, we evaluate two distinct demand forecasting solutions within a comprehensive, full-stack LEO satellite simulation compliant with DVB-S2X standards. Beyond prediction accuracy, we examine how incorporating user demand forecasts into BH plan generation impacts key network metrics, particularly delay and jitter. We evaluate these forecasting solutions alongside a static allocation baseline. Our results demonstrate that forecast-based dynamic planning reduces delay by 10-40\% across the beams under certain load conditions compared to static allocation methods. Crucially, marginal improvements in predictive accuracy do not translate into proportional network metric gains. While the evaluated forecasting solutions differ by 14-16\% in Normalized Mean Square Error (NMSE), this discrepancy yields less than a 1\% reduction in delay and produces nearly identical jitter characteristics. These findings suggest that when designing user demand forecasting solutions for practical LEO deployments, prioritizing system scalability may be more valuable than chasing minor accuracy enhancements.
\end{abstract}


\section{Introduction}
LEO constellations are expected to play a central role in the upcoming wireless ecosystem by complementing terrestrial networks and enabling seamless worldwide coverage \cite{azari2022evolution}. Crucially, next-generation networks must deliver reliable, high-capacity connectivity to remote, sparsely populated, or geographically challenging regions. However, deploying traditional terrestrial infrastructure in these environments is often economically unfeasible or physically impossible. Consequently, advanced satellite communication systems have emerged not just as a fallback, but as a primary enabler of next-generation wireless networks, bridging the digital divide to realize ubiquitous global connectivity for both civil \cite{kodheli} and military applications \cite{mil1}, \cite{mil2}.

A critical enabler for LEO satellites to deliver broadband connectivity is beam hopping (BH) technology. BH provides spatial and temporal flexibility in beam activation, allowing the satellite to dynamically match service provision with geographically fluctuating user demand \cite{yahia}. Consequently, radio resources are utilized much more efficiently compared to conventional continuous beam illumination. However, this flexibility introduces the challenges of frequent radio resource management. Depending on the configuration, the network must continuously determine the hopping sequence, the target cells, and the duration of each beam's illumination, as known Dwell Time (DT). Crucially, because these beam hopping plans must be generated and committed well in advance of their actual execution, there is a fundamental need to accurately forecast near-future user demand during the planning phase.

Motivated by this need, a growing body of literature has explored user demand forecasting in LEO satellite networks. For instance, recent studies have proposed various deep learning architectures, such as transformer-based models \cite{demirci1}, \cite{satform} and other form of neural network-based approaches \cite{bie2019combined} to predict user traffic fluctuations. However, all of these works typically isolate the forecasting mechanism, evaluating the proposed models solely on accuracy metrics. Consequently, they overlook the practical impact of these forecasts on the broader network layers. Improvements in predictive accuracy are only valuable if they translate into tangible enhancements in Quality of Service (QoS) parameters, such as reduced delay. Addressing this gap, we shift our focus from isolated statistical accuracy to network layer-wide performance by evaluating forecasting solutions within a comprehensive, end-to-end LEO satellite network simulation.

To construct a realistic protocol stack and accurately measure delay, we base our network architecture on Digital Video Broadcasting (DVB) standards. Although originally designed for Geostationary Earth Orbit (GEO) satellites employing continuous beam allocation, the DVB framework has been formally extended to support dynamic beam hopping \cite{dvbs2x}. Specifically, for forward link (downlink), the DVB-S2X standard has the notion of Beam Hopping Time Plans (BHTP), which dictate both the sequence of cell illumination and the duration of each beam's activation, known as the Dwell Time (DT). While a BHTP can be configured with a fixed plans, the standard also supports traffic-driven dynamic plans, an approach we adopt in this work to link user demand forecasts to actual resource allocation.

 \begin{figure*}[t!]
\centering
\includegraphics[width=\textwidth]{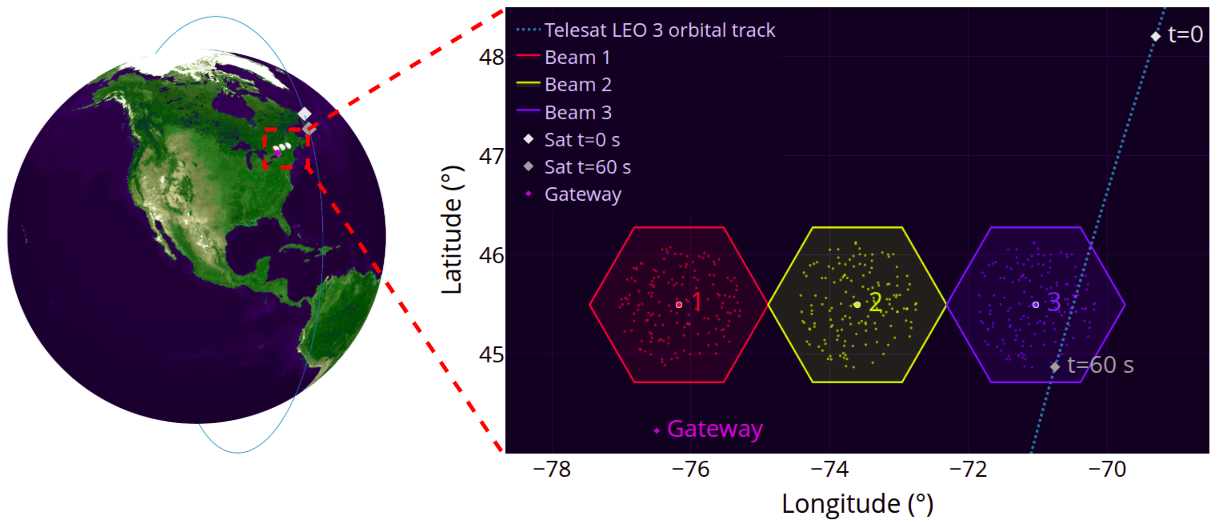} 
\vspace{-15pt}
\caption{Orbital trajectory of the Telesat LEO 3 satellite, generated using Two-Line Element (TLE) data. This one-second snapshot (19:07:30–19:07:31 UTC on April 16, 2026) captures the satellite's transit over Quebec. For the sake of simplicity, the satellite offers only one active beam at any given time. This single beam serves three cells on the forward link, establishing the baseline scenario used to investigate the impact of forecasting accuracy on further QoS metrics.}
\label{fig:taus}
\end{figure*}

More particularly we make the following contributions:
\begin{enumerate}[label=C\arabic*]
\item We establish a cross-layer evaluation methodology by integrating two demand forecasting solutions into an open source DVB-S2X simulation framework. By employing a traffic-driven beam hopping scheduler, we dynamically generate BHTPs based on anticipated user demand to investigate network-layer impacts.

\item We quantify the baseline benefits of dynamic resource allocation, demonstrating that forecast-based BHTPs strictly outperform static allocations by reducing end-to-end packet delay by 10\% to 40\% across diverse load conditions for different beams.

\item We reveal a non-proportional relationship between forecasting accuracy and end-user QoS. Specifically, we find that a substantial 14-16\% improvement in Normalized MSE (NMSE) translates to less than a 1\% improvement in delay and yields nearly identical jitter. This highlights a diminishing return on predictive accuracy, suggesting that future research on forecasting solutions may find greater practical value in focusing on system scalability and low computational overhead rather than strictly minimizing error metrics.
\end{enumerate}

The remainder of this paper is organized as follows. Section II details the system architecture and proposed methodology. Section III presents the numerical results and performance evaluation. Finally, Section IV concludes the paper.

\section{Methodology}
In this work, we focus on the forward link (downlink). We consider there are Inter-Satellite Link (ISL) connections between the satellites and that the satellites have a regenerative payload. For simplicity, we assume that each beam is associated with a single Earth-fixed cell and that only one beam is active at a time with a fixed DT. 

Given this setting, the user demand is aggregated at the cell level from the satellite's perspective. Furthermore, we assume the underlying traffic is predominantly IP-based, and demand fluctuates over short intervals with bursty patterns. Therefore, we follow a self-similar traffic model for the considered setting \cite{demirci1}.

\subsection{Traffic model}
\label{chap:traffic}
Let $\boldsymbol{A}_t$ represent the aggregate packet demand for a cell at time $t$: 
\begin{equation}
\boldsymbol{A}_t = mt + \sqrt{\alpha m}\boldsymbol{Z}_t, \text{ } t \in (-\infty, \infty)
\label{eq:selfsimilarProcess}
\end{equation}
where $m>0$ is a mean input rate, $\alpha>0$ is a variance coefficient and $\boldsymbol{Z}_t$ is the normalized fractional Brownian motion (fBm) with a Hurst parameter of $H \in [\frac{1}{2},1)$ \cite{norros2}. $\boldsymbol{Z}_t$ meets the following properties \cite{norros2}:
\begin{enumerate}[label=(\roman*)]
    \item $\boldsymbol{Z}$ has stationary increments \label{cond:1}.
    \item $\boldsymbol{Z}_0=0$ and $\mathbb{E}[\boldsymbol{Z}_t]=0$ for $\forall t$ \label{cond:2}.
    \item $\mathbb{E}[\boldsymbol{Z}_t]^2=|t|^{2H}$.\label{cond:3}
    \item $\boldsymbol{Z}$ has continuous paths.
    \item Finite-dimensional distributions of $\boldsymbol{Z}$ are Gaussian distributions. \label{cond:5}
\end{enumerate}

Under this traffic model, we consider two forecasting methods to serve as strong baselines for accuracy comparison. The first is the Fractional Autoregressive Integrated Moving Average (FARIMA) model. FARIMA processes are asymptotically second-order self-similar with a self-similarity parameter $d+1/2$, where $0<d<1/2$ \cite{leland2002self}. The second baseline is the method derived by Norros and Gripenberg, which provides an optimal forecast for the considered traffic model\cite{norrosForecast}.

\subsection{Forecasting solutions}
\subsubsection{FARIMA}
The FARIMA process generalizes the standard Autoregressive Integrated Moving Average (ARIMA) model by allowing the differencing parameter $d$ to take non-integer values. Adopting the notation introduced by Box and Jenkins \cite{beran2013long}, a FARIMA process $W_t$ is formally defined by the equation:
\begin{equation}
    \phi(B)(1-B)^d W_t = \psi(B)\varepsilon_t, \quad t \geq 1
    \label{eq:farima_main}
\end{equation}
where $\varepsilon_t$ is assumed to be independently and identically distributed (i.i.d.). with zero mean and finite variance $\sigma_{\varepsilon}^2$. The term $(1-B)^d$ represents the fractional differencing operator. For $d \in (-\frac{1}{2}, \frac{1}{2})$, this operator is defined via the binomial series expansion:
\begin{equation}
    (1-B)^d = \sum_{j=0}^\infty a_j B^j
    \label{eq:infSeries}
\end{equation}
where the coefficients $a_j$ are determined using the Gamma function $\Gamma(\cdot)$:
\begin{equation}
    a_j = \frac{\Gamma(j-d)}{\Gamma(j+1)\Gamma(-d)} = \frac{\Gamma(d+1)}{\Gamma(j+1)\Gamma(d-j+1)}(-1)^j
\end{equation}
The stationarity and invertibility of the process are determined by the autoregressive polynomial $\phi(k)=1-\sum_{j=1}^p\phi_jk^j$ and the moving-average polynomial $\psi(k)=\sum_{j=0}^q\psi_jk^j$. \cite{beran2013long}. 

\subsubsection{The Optimal Predictor for fBm}
Gripenberg and Norros \cite{norrosForecast} derived a closed-form optimal predictor for fBm driven processes. Let $\boldsymbol{Z}$ be an fBm process, as defined in Section \ref{chap:traffic}. For a forecasting horizon $\delta > 0$ and a look-back window $\tau \in (0, \infty]$, this optimal predictor is given by:
\begin{equation}
    \boldsymbol{\hat{Z}}_{\delta,\tau} = \mathbb{E}[\boldsymbol{Z}_\delta|\boldsymbol{Z}_s, s \in (-\tau,0)]
\label{eq:norFor}
\end{equation}
Eq.\eqref{eq:norFor} can be written as:
\begin{multline}
    \boldsymbol{\hat{Z}}_{\delta,\tau} = \int_{-\tau}^0 g_\tau(\delta,t) d \boldsymbol{Z}_t \quad \text{for }\tau<\infty, t\in(0,\tau)
\label{eq:forecaster}
\end{multline}
where
\begin{multline}
    g_\tau(\delta,-t) = \frac{sin(\pi(H-\frac{1}{2}))}{\pi}t^{-H+\frac{1}{2}}(\tau-t)^{(-H+\frac{1}{2})}\\\int_0^\delta\frac{z^{H-\frac{1}{2}} (z+\tau)^{H-\frac{1}{2}}}{z+t}dz.
\label{eq:smoothFunc}
\end{multline}
In the remainder of the paper, we refer to this optimal forecaster as ``Norros" forecaster.

\begin{algorithm}[t]
\caption{Dynamic Beam Hopping Plan Generation (A single active beam)}
\label{alg:dynamic_bh}
\small
\begin{algorithmic}[1]

\Statex \textbf{Input:} Per-cell traffic history $\{\mathbf{x}_c\}_{c \in \mathcal{C}}$,
                         where $\mathbf{x}_c = [\boldsymbol{A}^c[i-T+1], \ldots, \boldsymbol{A}^c[i]]$
                         are the last $T$ arrival observations (bytes per plan period)
\Statex \textbf{Output:} Beam hopping plan $\boldsymbol{\pi} = [\pi_1, \pi_2, \ldots, \pi_P]$,
                         an ordered sequence of $P$ cell IDs

\Statex
\Statex \textbf{Parameters:} $P$ = Slots per period (plan length),
                               $T$ = Historical look-back window length,
                               $\mathcal{C}$ = Set of cells,
                               $n = |\mathcal{C}|$, $\Delta$ = Forecast look-ahead horizon 

\Statex
\hrulefill
\Statex \textsc{Sub-step 1: Traffic Observation and History Update}
\hrulefill

\For{each cell $c \in \mathcal{C}$}
    \State $\boldsymbol{A}^c[i] \leftarrow \textsc{ReadAndReset}(\text{LLC}_c)$
    \State Append $\boldsymbol{A}^c[i]$ to $\mathbf{x}_c$; discard oldest sample if $|\mathbf{x}_c| > T$
\EndFor

\Statex
\hrulefill
\Statex \textsc{Sub-step 2: Demand Forecast and Mapping}
\hrulefill

\For{each cell $c \in \mathcal{C}$}
    \State $\hat{x}_c \leftarrow \textsc{Forecast}(c,\, \mathbf{x}_c)$\Comment{FARIMA or Norros}
    \State $s_c \leftarrow \textsc{ComputeShares}(\hat{x}_c)$ \Comment{Map forecast to slot demand}
\EndFor


\Statex
\hrulefill
\Statex \textsc{Sub-step 3: Plan Construction}
\hrulefill

\Statex \textit{Bresenham spread scheduling:}
\State $k_c \leftarrow 0$ for all $c$; \quad $\rho_c \leftarrow s_c$ for all $c$; \quad $\boldsymbol{\pi} \leftarrow []$
\For{$j = 1$ \textbf{to} $P$}
    \For{each cell $c$ with $\rho_c > 0$}
        \State $k_c \leftarrow k_c + s_c$ \Comment{Accumulate credit}
    \EndFor
    \State $c^* \leftarrow \arg\max_{c:\, \rho_c > 0}\; k_c$ \Comment{Select highest credit}
    \State Append $c^*$ to $\boldsymbol{\pi}$; \quad $\rho_{c^*} \leftarrow \rho_{c^*} - 1$; \quad $k_{c^*} \leftarrow k_{c^*} - P$
\EndFor

\State \Return $\boldsymbol{\pi}$

\end{algorithmic}
\end{algorithm}

\subsection{Dynamic scheduling, linking the forecasts to BH plans}

To facilitate implementation, we adopt a discrete-time model where each index represents a single forecasting period, the interval between successive plan-generation epochs. Let $P$ denote the duration of this period, linking continuous time $t$ to discrete index $i \in \mathbb{Z}$ via $t = iP$. Let $\mathcal{C}$ denote the set of available cells, where $A^c[i]$ denote the total traffic (in bytes) arriving at the Logical Link Control (LLC) layer of the satellite from the UTs located in cell $c$ during the $i$-th period, which spans the continuous-time interval $((i-1)P, iP]$.

To keep track of $A^c[i]$, we instrument a per-cell byte counter at the satellite. At the end of each forecasting period, the counter is read and reset to zero to yield the arriving demand for that cell. These periodic samples are accumulated into a sliding look-back window of length $T$, which serves as the input vector for the forecast models. Operating on these $T$ recent samples, the forecaster predicts the expected demand over the next discrete step, $\Delta$. 

Because the forecasting period is longer than the DT, each plan horizon spans $S$ consecutive slots. Thus, a 15 ms forecasting period combined with a 1 ms DT yields $S = 15$ slots. Based on the demand forecasts, the $S$ available slots are allocated across all beams for the upcoming plan period. To translate this allocation into a beam-hopping plan, we utilize Bresenham spreading scheduling \cite{bresen}. We chose a Bresenham-based algorithm over basic round-robin scheduling to interleave the slots in a way that maximizes the regularity of a cell's illumination. Algorithm \ref{alg:dynamic_bh} outlines the overall implementation.

\section{Numerical results}

\begin{table}[b]
\centering
\caption{Per-beam user terminal counts and Pareto shape parameters, and scheduling methods per experiment.}
\small 
\begin{tabular}{|c|l|c|cc|cc|cc|}
\hline
\multirow{2}{*}{Exp.} & \multirow{2}{*}{Method} & \multirow{2}{*}{$S$}
  & \multicolumn{2}{c|}{Beam 1} & \multicolumn{2}{c|}{Beam 2} & \multicolumn{2}{c|}{Beam 3} \\
\cline{4-9}
 & & & $\alpha$ & $M$ & $\alpha$ & $M$ & $\alpha$ & $M$ \\
\hline
1  & Static & \multirow{6}{*}{15} & 1.04 & 150 & 1.04 & 150 & 1.04 & 150 \\
2  & FAR. &                     & 1.04 & 150 & 1.04 & 150 & 1.04 & 150 \\
3  & Norros &                     & 1.04 & 150 & 1.04 & 150 & 1.04 & 150 \\
\cline{1-2}\cline{4-9}
4  & Static &                     & 1.04 & 210 & 1.24 & 150 & 1.44 &  90 \\
5  & FAR. &                     & 1.04 & 210 & 1.24 & 150 & 1.44 &  90 \\
6  & Norros &                     & 1.04 & 210 & 1.24 & 150 & 1.44 &  90 \\
\hline
\hspace{2pt}7 & Static &  & 1.04 & 150 & 1.04 & 150 & 1.04 & 150 \\
8  & FAR. & \multirow{4}{*}{45} & 1.04 & 150 & 1.04 & 150 & 1.04 & 150 \\
9  & Norros &                     & 1.04 & 150 & 1.04 & 150 & 1.04 & 150 \\
\cline{1-2}\cline{4-9}
\hspace{2pt}10  & Static &                     & 1.04 & 210 & 1.24 & 150 & 1.44 &  90 \\
11  & FAR. &                     & 1.04 & 210 & 1.24 & 150 & 1.44 &  90 \\
12 & Norros &                     & 1.04 & 210 & 1.24 & 150 & 1.44 &  90 \\
\hline
\end{tabular}
\label{table:exps}
\end{table}

To evaluate the proposed system, we simulated the considered setting using the sns-3 satellite extension \cite{sns3} for the ns-3 network simulator. Because the publicly available version of sns-3 does not natively support dynamic allocation, we implemented Algorithm \ref{alg:dynamic_bh} utilizing the framework's regenerative network option. While the core sns-3 simulator is written in C++, our forecasting algorithms are implemented in Python. We established a direct inter-process pipeline between the two environments at runtime. It is important to note that because ns-3 operates as a discrete-event simulator, the physical execution time of the Python forecasting scripts is decoupled from the simulated network clock; therefore, the computational delay of the forecasts is not reflected in the obtained end-to-end network delay values. Some of our changes are provided in the project's GitHub repository, yet one needs the ns-3 simulator alongisde the sns-3 extension to run these changes. \footnote{\url{https://github.com/YektaDemirci/forecastDVBperspective}}

To synthetically generate self-similar traffic, we employed the superposition of $M$ independent packet trains \cite{taqqu1997proof} per cell using the \textit{OnOffApplication} of ns-3, assigning the same pareto shape values for both ON and OFF periods. During an ON period, each UT demands 1Mbps of traffic. Given this superposition, there is the following relationship between the pareto shape $\alpha$ and the Hurst parameter $H$: $H=(3-\alpha)/2$ \cite{taqqu1997proof}.

To evaluate different network dynamics, we conducted the experiments denoted by ``Exp.” in Table \ref{table:exps}. We run each experiments 20 times with different seeds. We consider three allocation methods: one static and two dynamic, the latter based on forecasts from FARIMA (denoted as ``FAR") and Norros forecaster.

\begin{table}[b]
\centering
\caption{Physical Link Parameters}
\small
\centering
\begin{tabular}{|l|l|}
\hline
\textbf{Parameter} & \textbf{Value} \\
\hline
Feeder link base frequency & 27.5 GHz \\
Feeder link channels & 5 \\
Feeder carrier bandwidth & 1 GHz each \\
User link base frequency & 19.7 GHz \\
User link channels & 1 \\
User carrier bandwidth & 330 MHz \\
Roll-off & 0.2\\
MODCOD & QPSK 1/2 \\
\hline
\end{tabular}
\label{table:systemParams}
\end{table}

For static allocation, beam weights are assigned proportionally to the number of UTs per beam; for example, Experiments 1 and 7 use equal weights of $(5, 5, 5)$, whereas Experiments 4 and 10 use $(7,5,3)$. Meanwhile, the forecasters use the last 24 samples to forecast ($T=24$) and look-ahead horizon is 1 ($\Delta=1$).

Throughout these experiments, we assume a DT of 1 ms, such that the planning periodicity in milliseconds is numerically equal to the plan horizon in slots, denoted as $S$ in Table \ref{table:exps}. In Experiments 1, 2, and 3, utilizing a planning periodicity of 15 ms ($S = 15$), we simulate 150 User Terminals (UTs) per cell with uniform $\alpha$ values. For Experiments 4, 5, and 6, we introduce an asymmetric distribution of users; specifically, we assign a lower Pareto shape parameter $\alpha$, (higher Hurst), to the cell with the higher number of UTs to model different traffic load characteristics. Experiments 7–9 share the exact traffic generation setup as Experiments 1–3, but the planning periodicity is increased from 15 ms to 45 ms ($S = 45$). Similarly, Experiments 10–12 mirror Experiments 4–6, except for the extended 45 ms ($S = 45$) periodicity.

The Adaptive Coding and Modulation (ACM) configuration was fixed at QPSK 1/2 to investigate the direct impact of demand forecasting independent of channel impairments. Sufficient bandwidth was provisioned for the forward feeder link (GW-SAT), whereas the forward user link bandwidth was explicitly set to 330 MHz to closely match user demand load. Further details regarding the simulation setup of physical link are provided in Table  \ref{table:systemParams}.

\begin{figure*}[t]
\centerline{\includegraphics[width=\textwidth]{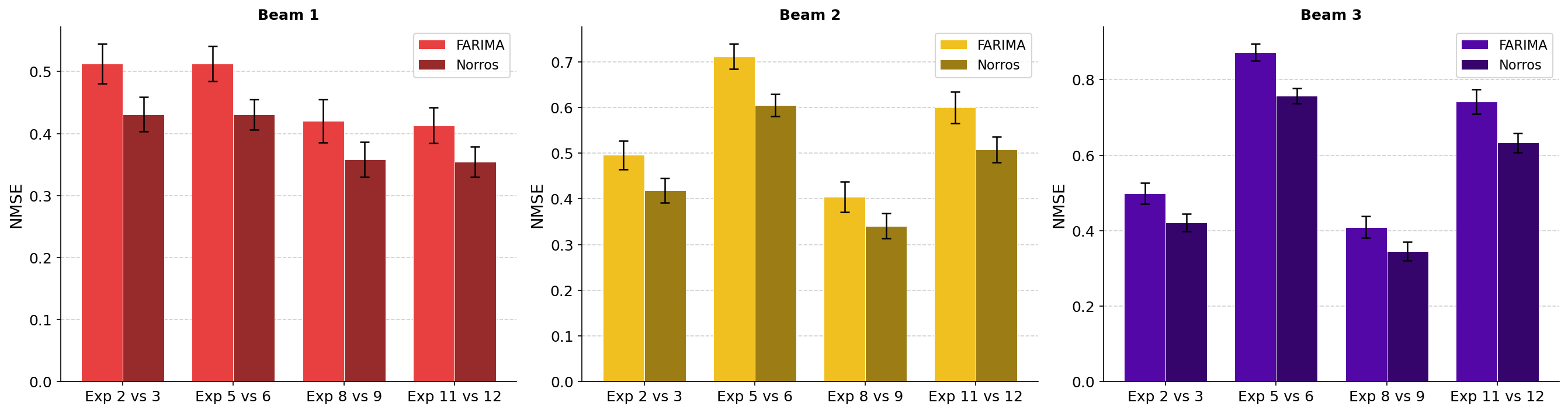}}
\caption{Comparison of NMSE between FARIMA and Norros forecasting models.}
\label{fig:nmse}
\end{figure*}

\subsection{Forecasting Accuracy Evaluation via NMSE}
Given these experiments, we first compare the forecasting accuracy of the considered forecasting solutions. We use NMSE, defined as:$$NMSE = \frac{\sum_{i=1}^n (y_i - \hat{y}_i)^2}{\sum_{i=1}^n (y_i - \bar{y})^2},$$ where $y_i$ is the true value, $\hat{y}_i$ is the forecasted value, and $\bar{y}$ is the sample mean of the observed data.

In Fig~\ref{fig:nmse}, the mean NMSE accuracy alongisde 95\% confidence interval is presented from 20 runs with different seeds. In all four pairs and on every beam, Norros forecaster yields a strictly lower NMSE than FARIMA. Considering the magnitude of improvement: Norros forecaster reduces NMSE by roughly 14-16\% relative to FARIMA.

For both forecasting solutions, the forecasting accuracy is directly affected by the burstiness of the underlying traffic. As we adopt lower $\alpha$ values (such as beam 1 in any experiment), the forecasting accuracy increases compared to higher $\alpha$ values (such as beam 3 from Experiments 4-6, 10-12).

\subsection{Delay measurements}

\begin{figure*}[b]
\centerline{\includegraphics[width=\textwidth]{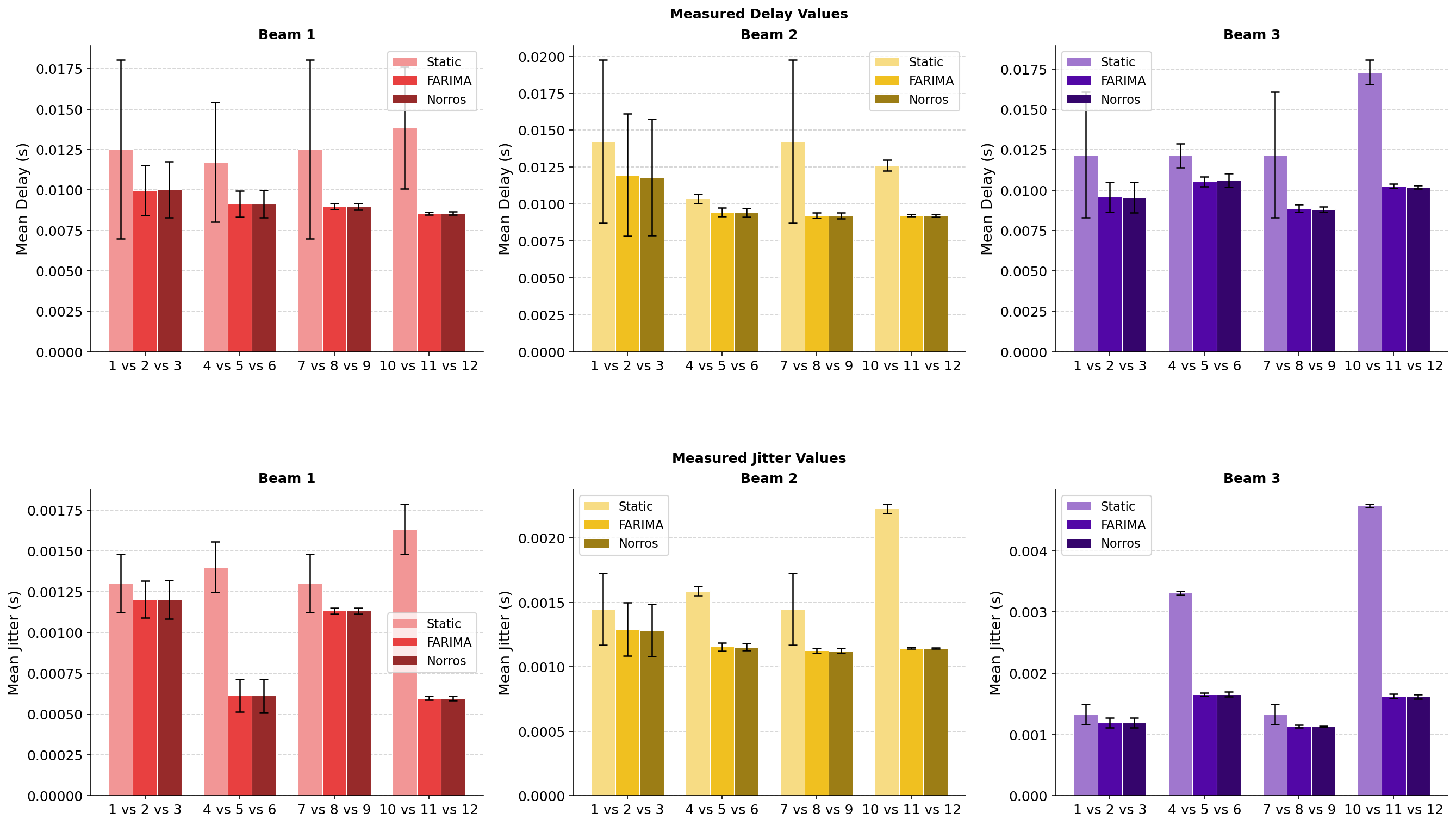}}
\caption{Comparison of delay and jitter performance between static allocation and dynamic allocations based on Norros and FARIMA models.}
\label{fig:delay}
\end{figure*}

Fig.~\ref{fig:delay} presents the one-way end-to-end delay at the LLC layer between the GW and each UT.

Between the two forecasting schemes, the Norros model demonstrates slightly better overall performance, achieving the lowest delay in 8 of the 12 evaluated beam-experiment combinations. However, it is difficult to declare it the definitively superior model. The two frameworks perform virtually head-to-head, with performance gaps consistently averaging less than 1\%. On the other hand, both dynamic forecasting models yield substantial improvements over the static baseline across all tested scenarios.

Under a 15ms planning periodicity, the models trade marginal gains depending on the load and the specific beam. During symmetric loading, Norros (Exp 3) edges out FARIMA (Exp 2) on Beams 2 and 3 by 1.3\% and 0.2\% respectively, while FARIMA counters with a 0.5\% lead on Beam 1. A similar back-and-forth occurs under asymmetric loading: FARIMA (Exp 5) takes Beams 1 and 3 by narrow margins of 0.1\% and 0.8\%, while Norros (Exp 6) claims Beam 2 by 0.3\%. Despite these fractional differences between the two, both models successfully reduce delay by 17\% to 20\% in the symmetric tests and 10\% to 22\% in the asymmetric tests compared to their respective static baselines.

When the planning periodicity shifts to 45ms, Norros establishes a slightly firmer lead. Under symmetric loading, it outperforms FARIMA across all three beams by margins ranging from 0.1\% to 0.8\%. In the asymmetric scenario, Norros (Exp 12) continues this trend on Beams 2 and 3 with improvements of 0.2\% and 0.7\%, though FARIMA (Exp 11) manages to edge ahead on Beam 1 by a mere 0.2\%. This longer planning periodicity further highlights the strength of both forecasting models against the static baselines, driving massive delay reductions of 28\% to 35\% for symmetric loads and 25\% to 40\% for asymmetric loads.

\subsection{Jitter measurements}
Jitter is measured over the same path as delay, from the LLC layer of the GW to the LLC layer of the UTs.

Under a 15ms planning periodicity, the two models deliver almost identical jitter performance. The jitter differences are too small to be practically significant. Despite producing these almost identical results, both dynamic models successfully reduce jitter by 7\% to 11\% in symmetric tests and a substantial 27\% to 56\% in asymmetric tests compared to the static baselines.

When the planning periodicity shifts to 45ms, Norros establishes a slightly firmer lead on Beams 2 and 3 under both loading conditions, but their outputs remain almost identical, even perfectly tying on Beam 1. Again, it is difficult to declare a definitively superior model. 45ms periodicity highlights the strength of dynamic allocation against the static baselines, driving jitter reductions of 13\% to 22\% for symmetric loads and 48\% to 65\% for asymmetric loads.

\section{Conclusion}

To investigate the impact of user demand forecasting on higher-layer network parameters, we conducted an end-to-end LEO satellite simulation compliant with DVB-S2X standards. We evaluated two forecasting solutions, FARIMA and an optimum forecaster, employing a single-beam setup. Using a self-similar traffic model, we demonstrate that forecast-based resource allocation in LEO satellite beam-hopping systems significantly outperforms static scheduling, reducing end-to-end packet delay by 10\% to 40\%.

While this performance gain is expected, our most notable finding is the disconnect between standalone predictive accuracy and end-user QoS. Although the evaluated forecasting models exhibited an 8\% to 16\% improvement in Normalized Mean Square Error (NMSE), this translated to less than a 1\% difference in network delay and nearly identical jitter. These results suggest that future LEO deployments should prioritize forecasting solutions that offer high system scalability and low computational overhead, rather than pursuing marginal improvements in forecasting accuracy metrics. 

Future work will extend this analysis to the constellation level to investigate how ISL influence end-to-end delay and jitter under dynamic beam hopping. Additionally, we plan to explore how these forecasting frameworks perform under highly heterogeneous traffic profiles, ensuring resilient QoS for diverse applications.

\section*{Acknowledgment}
This work was supported in part by MDA Space; in part by the Consortium de Recherche et d’innovation en Aérospatiale au Québec (CRIAQ); and in part by the Natural Sciences and Engineering Research Council of Canada (NSERC).

\bibliographystyle{IEEEtran} 
\bibliography{main}

@book{beran2013long,
  title={Long-Memory Processes: Probabilistic Properties and Statistical Methods},
  author={Beran, J. and Feng, Y. and Ghosh, S. and Kulik, R.},
  isbn={9783642355127},
  series={SpringerLink : B{\"u}cher},
  year={2013},
  publisher={Springer Berlin Heidelberg}
}

@article{norrosForecast,
  title={On the prediction of fractional Brownian motion},
  author={Gripenberg, Gustaf and Norros, Ilkka},
  journal={Journal of Applied Probability},
  volume={33},
  number={2},
  pages={400--410},
  year={1996},
  publisher={Cambridge University Press}
}

@ARTICLE{leland2002self,
  author={Leland, W.E. and Taqqu, M.S. and Willinger, W. and Wilson, D.V.},
  journal={IEEE/ACM Transactions on Networking}, 
  title={On the self-similar nature of {E}thernet traffic (extended version)}, 
  year={1994},
  volume={2},
  number={1},
  pages={1-15},
  doi={10.1109/90.282603}}

@ARTICLE{norros2,
  author={Norros, I.},
  journal={IEEE Journal on Selected Areas in Communications}, 
  title={On the use of fractional Brownian motion in the theory of connectionless networks}, 
  year={1995},
  volume={13},
  number={6},
  pages={953-962},
  doi={10.1109/49.400651}}

@techreport{bresen,
  author       = {Carl A. Waldspurger and William E. Weihl},
  title        = {Stride Scheduling: Deterministic Proportional-Share Resource Management},
  institution  = {Massachusetts Institute of Technology, Laboratory for Computer Science},
  type         = {Technical Memorandum},
  number       = {MIT/LCS/TM-528},
  address      = {Cambridge, MA, USA},
  year         = {1995},
  month        = jun
}

@inproceedings{sns3,
  title={Satellite model for network simulator 3.},
  author={Puttonen, Jani and Rantanen, Sami and Laakso, Frans and Kurjenniemi, Janne and Aho, Kari and A{\c{c}}ar, G{\"u}ray},
  booktitle={SimuTools},
  pages={86--91},
  year={2014}
}

@article{azari2022evolution,
  title={Evolution of non-terrestrial networks from 5{G} to 6{G}: A survey},
  author={Azari, M Mahdi and others},
  journal={IEEE communications surveys \& tutorials},
  volume={24},
  number={4},
  pages={2633--2672},
  year={2022},
  publisher={IEEE}
}

@inproceedings{mil1,
  title={Failure resilience in proliferated low Earth orbit satellite network topologies},
  author={Shake, Thomas and Sun, Jun and Royster, Thomas and Narula-Tam, Aradhana},
  booktitle={MILCOM 2022-2022 IEEE Military Communications Conference (MILCOM)},
  pages={828--834},
  year={2022},
  organization={IEEE}
}

@inproceedings{mil2,
  title={Fairness-aware scheduling optimization for NB-IoT in LEO satellite networks using a 3d spherical coordinate system},
  author={Lee, Byeongheon and Lee, Ju-Hyung and Ko, Young-Chai},
  booktitle={MILCOM 2023-2023 IEEE Military Communications Conference (MILCOM)},
  pages={957--962},
  year={2023},
  organization={IEEE}
}

@article{kodheli,
  title={Satellite communications in the new space era: A survey and future challenges},
  author={Kodheli, Oltjon and Lagunas, Eva and Maturo, Nicola and Sharma, Shree Krishna and Shankar, Bhavani and Montoya, Jesus Fabian Mendoza and Duncan, Juan Carlos Merlano and Spano, Danilo and Chatzinotas, Symeon and Kisseleff, Steven and others},
  journal={IEEE Communications Surveys \& Tutorials},
  volume={23},
  number={1},
  pages={70--109},
  year={2020},
  publisher={IEEE}
}

@article{yahia,
  title={Evolution of high-throughput satellite systems: A vision of programmable regenerative payload},
  author={Yahia, Olfa Ben and Garroussi, Zineb and B{\'e}langer, Olivier and Sans{\`o}, Brunilde and Frigon, Jean-Fran{\c{c}}ois and Martel, St{\'e}phane and Lesage-Landry, Antoine and Karabulut Kurt, Gunes},
  journal={IEEE Communications Surveys \& Tutorials},
  volume={27},
  number={3},
  pages={1565--1597},
  year={2024},
  publisher={IEEE}
}

@article{satform,
  title={Satformer: Accurate and robust traffic data estimation for satellite networks},
  author={Qin, Liang and Liu, Xiyuan and Wei, Wenting and Liang, Chengbin and Gu, Huaxi},
  journal={Advances in Neural Information Processing Systems},
  volume={37},
  pages={47530--47558},
  year={2024}
}

@article{demirci1,
  title={Forecasting Self-Similar User Traffic Demand Using Transformers in {LEO} Satellite Networks},
  author={Demirci, Yekta and Mantelet, Guillaume and Martel, St{\'e}phane and Frigon, Jean-Fran{\c{c}}ois and Karabulut Kurt, Gunes},
  journal={IEEE International Conference on Wireless for Space and Extreme Environments (WiSEE)},
  year={2025}
}

@misc{dvbs2x,
  author       = "{ETSI}",
  title        = "{EN 302 307-2 V1.4.1 (2024-08): Digital Video Broadcasting (DVB); Part 2: DVB-S2 Extensions (DVB-S2X)}",
  howpublished = "\url{https://www.etsi.org/deliver/etsi_en/302300_302399/30230702/01.04.01_60/en_30230702v010401p.pdf}",
  month        = aug,
  year         = 2024,
}

@article{taqqu1997proof,
author = {Taqqu, Murad S. and Willinger, Walter and Sherman, Robert},
title = {Proof of a fundamental result in self-similar traffic modeling},
year = {1997},
issue_date = {Apr. 1997},
publisher = {Association for Computing Machinery},
address = {New York, NY, USA},
volume = {27},
number = {2},
issn = {0146-4833},
doi = {10.1145/263876.263879},
journal = {SIGCOMM Comput. Commun. Rev.},
month = apr,
pages = {5–23},
numpages = {19}
}

@ARTICLE{bie2019combined,
  author={Bie, Yuxia and Wang, Longzi and Tian, Ye and Hu, Zhi},
  journal={IEEE Access}, 
  title={A Combined Forecasting Model for Satellite Network Self-Similar Traffic}, 
  year={2019},
  volume={7},
  number={},
  pages={152004-152013},
  doi={10.1109/ACCESS.2019.2944895}}

\end{document}